\documentclass{aa}
\usepackage{txfonts}
\usepackage{graphicx}
\usepackage{natbib}
\usepackage{longtable,lscape}
\bibpunct{(}{)}{;}{a}{}{,}
\begin{document}
\title{Long-term radio variability of AGN: flare characteristics}

\author{T. Hovatta \inst{1} \and E. Nieppola \inst{1} \and M. Tornikoski \inst{1} \and E. Valtaoja \inst{2,3} \and M.F. Aller \inst{4} \and H.D. Aller \inst{4}}
\institute{Mets\"ahovi Radio Observatory, TKK, Helsinki University of Technology, Mets\"ahovintie 114, 02540 Kylm\"al\"a, Finland  \\ \email{tho@kurp.hut.fi} \and Tuorla Observatory, University of Turku, V\"ais\"al\"antie 20, 21500 Piikki\"o, Finland \and Department of Physics, University of Turku, 20100 Turku, Finland \and Department of Astronomy, University of Michigan, Ann Arbor, MI 48109, USA}

\date{Received / Accepted}
\abstract
{}
{We have studied the flare characteristics of 55 AGN at 8 different frequency bands between 4.8 and 230\,GHz. Our extensive database enables us to study the various observational properties of flares in these sources and compare our results with theoretical models.}
{We visually extracted 159 individual flares from the flux density curves and calculated different parameters, such as the peak flux density and duration, in all the frequency bands. The selection of flares is based on the 22 and 37\,GHz data from Mets\"ahovi Radio Observatory and 90 and 230\,GHz data from the SEST telescope. Additional lower frequency 4.8, 8, and 14.5\,GHz data are from the University of Michigan Radio Observatory. We also calculated variability indices and compared them with earlier studies.}
{The observations seem to adhere well to the shock model, but there is still large scatter in the data. Especially the time delays between different frequency bands are difficult to study due to the incomplete sampling of the higher frequencies. The average duration of the flares is 2.5 years at 22 and 37\,GHz, which shows that long-term monitoring is essential for understanding the typical behaviour in these sources. It also seems that the energy release in a flare is independent of the duration of the flare.}
{}
\maketitle
\section{Introduction}
Active galactic nuclei (AGNs) exhibit variability across the whole 
electromagnetic spectrum, and it is often 
possible to see distinct flares in their flux density curves. By studying the 
differences in the variability of the various AGN types we gain understanding 
into the physics involved in producing these events. The study of the radio 
regime is of special 
interest because it relates directly to the shock development in the jets. 

Variability studies of a large sample of AGNs at lower radio 
frequencies of 4.8--14.5\,GHz 
have been done at the University of Michigan Radio Observatory (UMRAO), where 
AGNs have been monitored since the 1960s. \cite{hughes92} studied the long-term 
variability time scales of quasars and BL Lacertae objects (BLOs).
\cite{aller99} studied the total flux density and linear polarisation properties of 
BLOs and \cite{aller03} the variability of the complete Pearson--Readhead 
radio sample.

At Mets\"ahovi Radio Observatory AGNs have been monitored at 22 and 37\,GHz 
for almost 30 years. Variability studies were first performed after 
5 years of observations \citep{valtaoja88, valtaoja92, valtaoja92b} and 
updated by \cite{lainela94} with 12 years of data. Long-term variability 
time scales were studied by \cite{lainela93} and \cite{hovatta07}.
Southern AGNs were studied at the higher frequencies of 90 and 230\,GHz by 
\cite{tornikoski93} and \cite{tornikoski00}. Also \cite{ciaramella04} 
studied the variability properties of a sample of AGNs using the 
data from UMRAO and Mets\"ahovi.

Apart from these studies, most of the work done on the variability of AGNs 
concentrates on individual sources and their multifrequency behaviour. 
For example, in \cite{pyatunina06} and \cite{pyatunina07} seven AGNs are 
studied at 4.8--37\,GHz. 
They decompose the flux density curves into Gaussian flares and study the 
properties of the flares and also the duty cycles in these sources.
While this can describe and model one source very well, studies of large 
samples are important so that statistical differences between the source 
types can be studied.

The flares in the radio regime are explained by adiabatic shock-in-jet 
models \citep{marscher85, hughes85}. In these models an adiabatic shock is 
formed somewhere in the base of the jet and as it moves downward in the jet, 
we see synchrotron radiation at different frequencies and the observed properties 
depend on the phase of the shock development. According to the model by 
\cite{marscher85} the shock goes through three phases: The growth stage where 
Compton losses dominate and the flux density increases at all frequencies, 
a plateau stage where the energy gains and losses are equal and the peak 
frequency moves to lower frequencies, and a decay 
phase where adiabatic losses start to dominate and the flux density declines.

\cite{valtaoja92} developed a parametrisation of the Marscher \& Gear model 
-- a generalised shock model -- where the observed 
properties depend on the peak frequency of the synchrotron flare and how it 
relates to the observing frequency. 
If the peak frequency $\nu_\mathrm{max}$, where 
$S(\nu_\mathrm{max})$ reaches its highest value, is 
higher than the observing frequency $\nu_\mathrm{obs}$, the flare is 
classified as a high-peaking flare. In this case the shock reaches its 
maximal development above the observing frequency and is in its decaying 
stage when observed. We should also see time delays towards lower frequencies.
The flare is called low-peaking if $\nu_\mathrm{obs} > \nu_\mathrm{max}$. Now 
the shock is in its growing stage when observed and flux density curves of nearby 
frequencies trail each other closely and there should not be any time delays. 

This model can be tested by studying individual flares in AGNs. In this paper we 
used the Mets\"ahovi monitoring sample as a starting point and 
extracted 159 individual flares from the flux density curves of 55 sources. We 
used 8 different frequency bands between 4.8 and 230\,GHz 
to study the flare development. 
We also conducted a basic variability analysis by calculating the variability 
indices.

The paper is organised as follows: In Sect. \ref{sec:data} we describe the 
source sample and data used. The methods used are described in Sect. 
\ref{sec:methods}. Section \ref{sec:prop} includes the results of our 
analysis and the discussion follows in Sect. \ref{sec:discussion}. The 
conclusions are drawn in Sect. \ref{sec:conclusions}. We use the value 
$H_0 = 72$~km~$\mathrm{s}^{-1}$~$\mathrm{Mpc}^{-1}$
throughout the paper. 
All the statistical analyses have been performed with the 
Unistat statistical package for Windows\footnote{http://www.unistat.com/} (version 5.0).

\section{Data and the sample}\label{sec:data}
Our sample is selected from the Mets\"ahovi monitoring sources 
\citep{salonen87, terasranta92, terasranta98, terasranta04, terasranta05} 
and from our monitoring sample of southern AGNs at the Swedish-ESO Submillimetre
telescope (SEST) \citep{tornikoski96}. The criterium for selection was 
that there 
must be at least one well-sampled outburst at two of the 
frequency bands between 22 and 
230\,GHz. This criterium was fulfilled for 55 sources.
The minimum relative flux density change in these outbursts was 0.5 Jy 
at 37\,GHz and between 0.4 and 0.6 Jy at other frequency bands.
In addition to the 
Mets\"ahovi 22, 37 and 90\,GHz and SEST 90 and 230\,GHz data we used 
the lower frequency data at 4.8, 8 and 14.5\,GHz from the UMRAO database. 
Higher frequency data at 90, 150 and 230\,GHz 
were also collected from the literature 
\citep{Steppe88, Steppe92, Steppe93, Reuter97}.  Some of the southern sources 
are not monitored at Mets\"ahovi and thus have data only at 
the higher 90 and 230\,GHz and the UMRAO low frequencies.

Many of the sources have been monitored for over 25 years and here we used 
data from the beginning of our monitoring in the 1980s until 2005. 
Our study also includes unpublished data at 37\,GHz from December 2001 
to April 2005 and at 90 and 230\,GHz from the SEST from 1994.5--2003. 
The 37\,GHz data for BL Lacertae objects from 2001--2005 
are published in \cite{nieppola07}. The observation method and data reduction 
process of Mets\"ahovi data is described in \cite{terasranta98}. 

Our sample consists of different types of AGNs, of which 14 sources are 
BLOs, 4 are radio galaxies (GALs), 20 are Highly 
Polarised Quasars (HPQs), 12 are Low Polarisation Quasars (LPQs) and for 
5 sources we had no information about their optical polarisation and they are 
therefore considered to be LPQs in this study. The BLOs are studied in more 
detail in a separate paper \citep{nieppola08}. When 
studying the differences between the source classes GALs are not considered 
because of their smaller number in the study. The source sample is shown in 
Table \ref{table:sourcelist}\addtocounter{table}{1}, where the B1950-name, other name, classification, possible EGRET detection,
number of flares, number of datapoints since 1980 in each frequency band, 
maximum observed flux density at 37\,GHz, variability index at 37\,GHz and 
the median duration of flares at 37\,GHz are tabulated. 
There is a clear selection effect in the sample as only sources with 
distinct flares were selected for the analysis. The radio properties 
of these sources are interesting in view of the GLAST-satellite mission and 
therefore in Table \ref{table:sourcelist} we have also marked the 29 sources that are 
high-confidence or possible detections of EGRET \citep{hartman99, mattox01}. 

\section{Methods}\label{sec:methods}
\begin{figure}
\resizebox{\hsize}{!}{\includegraphics{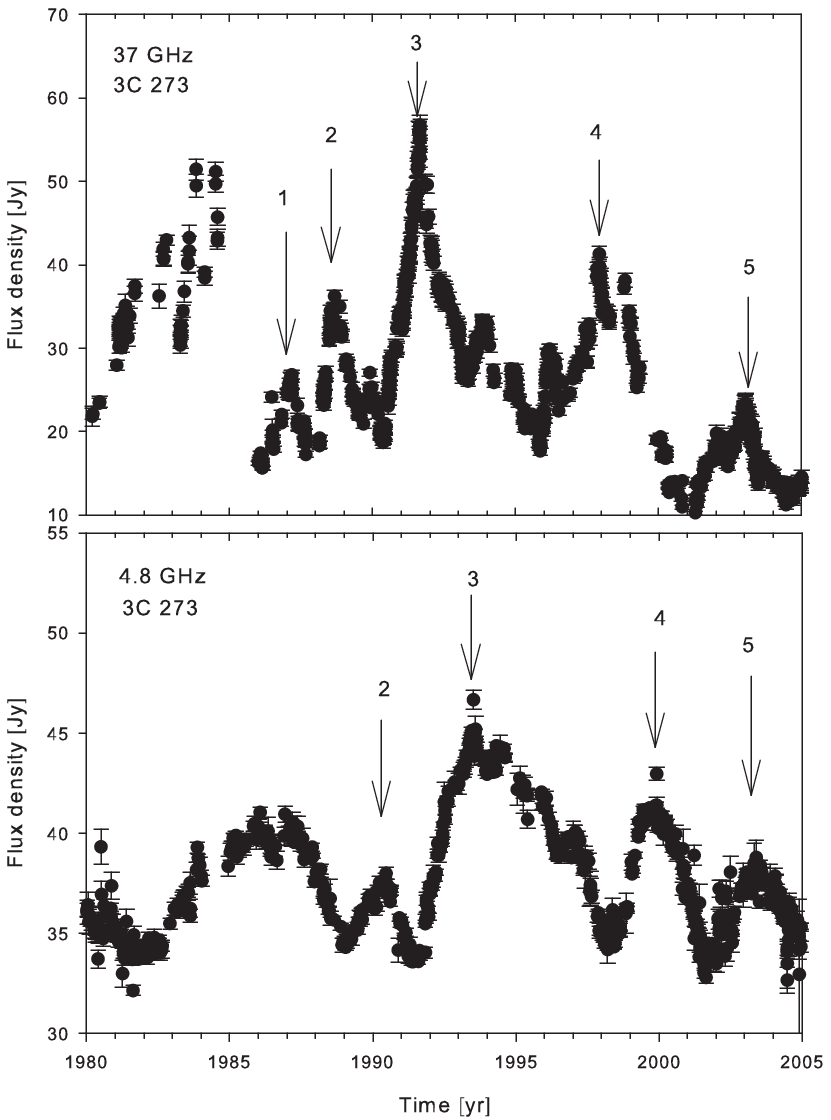}}
\caption{Upper panel: Flux density curve of the LPQ source \object{3C 273} (\object{1226+026}) at 37\,GHz. Flares used in our analysis are marked with arrows. Lower panel: Flux density curve at 4.8\,GHz}
\label{3C273}
\end{figure}

We separated each well-monitored flare within the flux density curves and 
determined their start, peak and end epochs and flux densities for those epochs.
Figures \ref{3C273} and \ref{0235} show examples of flux density curves. In 
Fig. \ref{3C273} the analysed flares of the LPQ source \object {3C 273} 
(\object{1226+023}) are shown at 37 and 4.8\,GHz.
Altogether 159 flares fulfilled our criterium of being well-monitored in at 
least two frequency bands. From the start and end times we calculated the 
duration of the flares in the observer's frame, and using the peak epoch, 
also the rise and decay 
times at all frequency bands separately. In addition to the direct 
observational peak flux density of the flare, we used in our analysis 
the relative peak flux density, which is the peak flux density minus the 
start flux density. Our flares are distributed between the 
classes so that there are 46 flares in BLOs, 15 in GALs, 59 in HPQs and 
39 in LPQs. This means that on average every BLO has 3.3 well-monitored 
flares during the 25 years while LPQs have one less (2.3). There are differences 
within the source classes as well.  For example, the 
BLO source \object{PKS 0735+178} 
has only one well-monitored flare during the 20 years' period it 
has been observed whereas \object{BL Lac} and \object{OJ 287} both have 9 
well-observed flares.

We note that our definition of individual flares is based on visual 
inspection and is clearly different compared to models where a theoretical 
flare form is fitted to the observations \citep[e.g.][]{valtaoja99, turler99}. 
Here we have, however, studied the observational properties of the 
flares and how they depend on each other. Therefore, a flare, as defined in this 
work, may include several individual shocks or events.

\section{Variability properties}\label{sec:prop}
\subsection{Variability indices}\label{sec:index}
\begin{figure*}
\includegraphics[width=17cm]{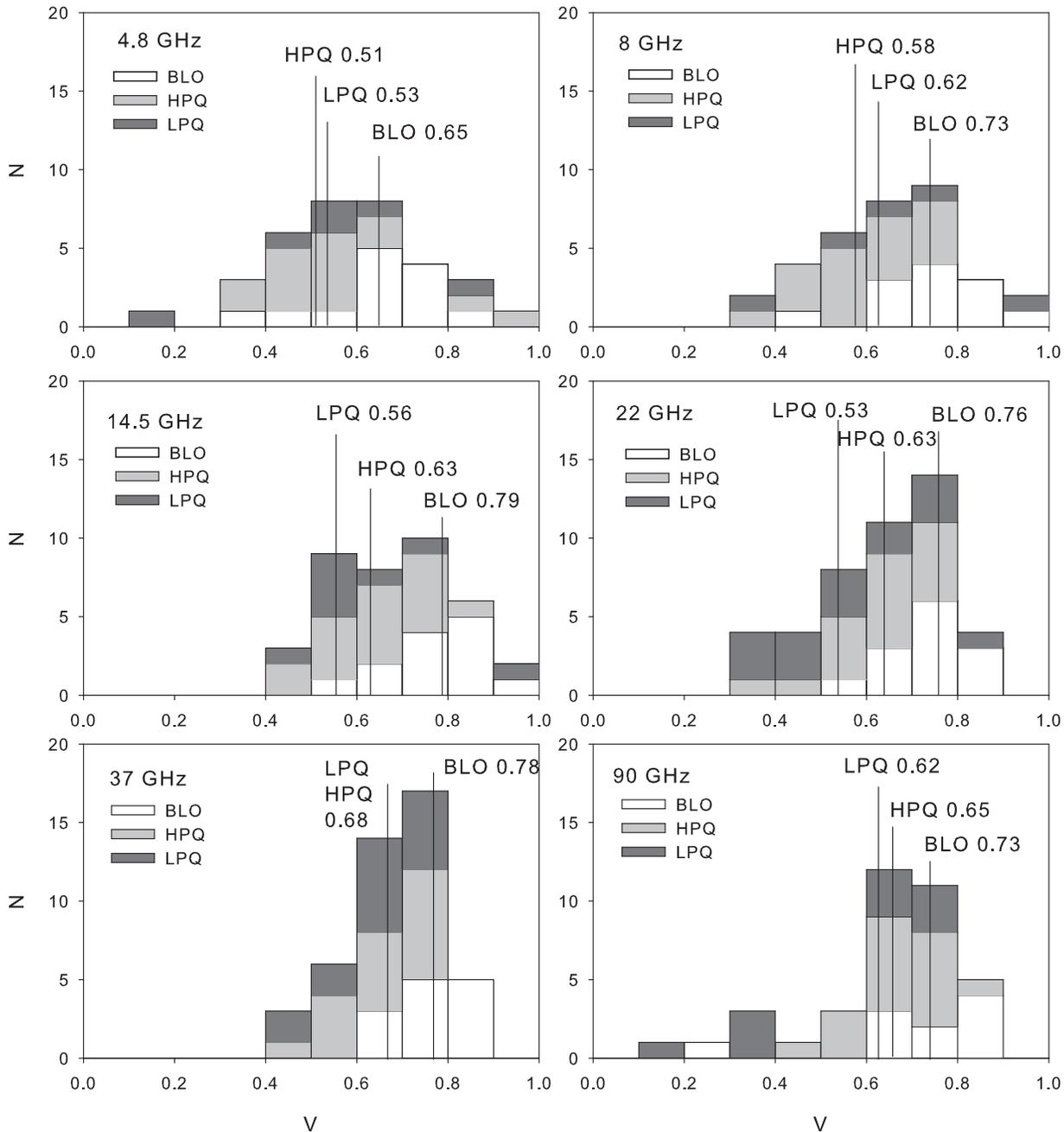}
\caption{Distribution of variability indices in different source classes and frequency bands. Median values for each source class are shown by vertical lines.}
\label{histo_var}
\end{figure*}

We calculated the variability indices $V$ for each source at all 
the frequency bands as defined in Eq. \ref{eq:varindex}
\begin{equation} \label{eq:varindex}
 V = \frac{(S_\mathrm{max} - \sigma_{S_\mathrm{max}}) - (S_\mathrm{min} + \sigma_{S_\mathrm{min}})}{(S_\mathrm{max}  - \sigma_{S_\mathrm{max}}) + (S_\mathrm{min} + \sigma_{S_\mathrm{min}})},
\end{equation}
where $S_\mathrm{max}$ is the maximum observed flux density of the source and 
$\sigma_{S_\mathrm{max}}$ is its error and $S_\mathrm{min}$ and 
$\sigma_{S_\mathrm{min}}$ are the minimum flux density and its error.
The distribution of the indices in the different source classes and at 
frequencies between 4.8 and 90\,GHz are shown in Fig. \ref{histo_var}, where 
also the median values are marked for each source class by vertical lines.

We ran the Kruskal-Wallis analysis (KW-analysis) to see if there 
are any statistically significant differences in the variability indices 
of the frequency bands and source classes. The only difference between 
the frequency bands was that the 4.8\,GHz
indices differ from most other frequencies significantly within 
the 95\% confidence 
limit. The indices at 4.8\,GHz are smaller than those at higher frequencies.
This can be understood as towards the lower frequency bands the flares are 
observed later due to synchrotron self-absorption. The flux densities are lower, 
timescales are longer, and it is also more difficult to distinguish the 
individual subsequent flares from each other. This is clearly seen in 
Fig. \ref{3C273} if the 4.8\,GHz flux density curve is compared to the 
37\,GHz curve. The relative flux densities are much lower and many individual flares 
seen in the 37\,GHz curve, for example the flare in 1996, cannot be separated 
in the 4.8\,GHz curve.

When studying the distributions it seems that the BLOs have larger variability 
indices than the quasars. This is also confirmed by the KW-analysis which 
gives statistically significant differences between BLOs and both quasar types at 
14.5--37\,GHz. At 4.8 and 8\,GHz LPQs and BLOs do not differ 
significantly but this is probably due to the low number of LPQs at these 
frequency bands. At 4.8\,GHz there are only 6 LPQs and at 8\,GHz only 5. 
These are also the more extreme LPQs with higher variability. In general, the 
individual flares of relatively faint LPQs are often superposed at lower 
frequencies.

\subsection{Flare amplitudes}\label{sec:parameters}
The number of flares for which it was possible to determine 
the peak flux densities 
vary between the frequency bands. Most of the flares 
are defined at 22 and 37\,GHz 
which were used for selecting the flares.
We determined the observed peak flux density for 153 flares at 22 and 37\,GHz 
whereas this was possible for only 33 flares at 230\,GHz. 
The number of sources for which we were able to define the relative flux density is 
even smaller because it was not always possible to determine when the flare 
started and what was the flux density then, even if the observed peak flux density is 
defined. In some cases we used the next minimum after the flare for 
the relative flux density. 

The observed peak flux densities and relative flux densities varied also
significantly between the flares. The minimum peak flux density observed was 0.7 Jy
and the maximum 56.9 Jy. Clearly there are always some extreme objects which make the 
scatter large but the median values of these parameters seem to describe the 
overall behaviour in each source class and frequency band quite well. 
The median values of observed peak and relative flux densities of the flares are 
shown in Table \ref{table:parameters} for each frequency band
and source class. 

\begin{table*}
\caption[]{Median values of duration, observed peak flux density and relative peak flux density for all frequencies and source classes. Number of flares used to calculate the value is also shown.}
\label{table:parameters}
\centering
\begin{tabular}{lcc|cc|cc|cc|cc}
\hline
\hline
$\nu$ [GHz]	  & \multicolumn{10}{c} {median duration [yr]}	  	  \\
 	  & 	 All 	  & 	 N 	  & 	 BLO 	  & 	 N 	  & 	 GAL  	  & 	 N 	  & 	 HPQ  	  & 	 N 	  & 	 LPQ 	  & 	 N 	  \\
\hline 
 4.8 	  & 	 2.9 	  & 	 81 	  & 	 2.7 	  & 	 34 	  & 	 2.8 	  & 	 7 	  & 	 3.0 	  & 	 30 	  & 	 3.6 	  & 	 10 	  \\
 8 	  & 	 2.8 	  & 	 94 	  & 	 2.7 	  & 	 37 	  & 	 2.6 	  & 	 8 	  & 	 2.8 	  & 	 37 	  & 	 3.3 	  & 	 12 	  \\
 14.5 	  & 	 2.5 	  & 	 110 	  & 	 2.4 	  & 	 44 	  & 	 2.2 	  & 	 8 	  & 	 2.5 	  & 	 40 	  & 	 3.1 	  & 	 18 	  \\
 22 	  & 	 2.5 	  & 	 145 	  & 	 2.3 	  & 	 43 	  & 	 2.4 	  & 	 14 	  & 	 2.5 	  & 	 53 	  & 	 2.6 	  & 	 35 	  \\
 37 	  & 	 2.4 	  & 	 150 	  & 	 2.4 	  & 	 43 	  & 	 2.1 	  & 	 15 	  & 	 2.4 	  & 	 56 	  & 	 2.7 	  & 	 36 	  \\
 90 	  & 	 2.3 	  & 	 68 	  & 	 2.5 	  & 	 18 	  & 	 2.1 	  & 	 3 	  & 	 2.1 	  & 	 31 	  & 	 3.0 	  & 	 16 	  \\
 230 	  & 	 2.5 	  & 	 31 	  & 	 2.3 	  & 	 9 	  & 	 $-$ 	  & 	 $-$ 	  & 	 2.8 	  & 	 15 	  & 	 2.5 	  & 	 7 	  \\
\hline 
\\
\hline
 $\nu$ [GHz]	  & \multicolumn{10}{c} {median observed peak flux density [Jy]} 	  \\
	  & 	 All 	  & 	 N 	  & 	 BLO 	  & 	 N 	  & 	 GAL  	  & 	 N 	  & 	 HPQ  	  & 	 N 	  & 	 LPQ 	  & 	 N 	  \\
\hline 
 4.8 	  & 	 4.0 	  & 	 85 	  & 	 3.2 	  & 	 34 	  & 	 5.0 	  & 	 8 	  & 	 3.7 	  & 	 33 	  & 	 8.1 	  & 	 10 	  \\
 8 	  & 	 4.6 	  & 	 99 	  & 	 4.3 	  & 	 38 	  & 	 3.9 	  & 	 8 	  & 	 4.5 	  & 	 41 	  & 	 10.2 	  & 	 12 	  \\
 14.5 	  & 	 5.0 	  & 	 117 	  & 	 4.2 	  & 	 45 	  & 	 3.6 	  & 	 9 	  & 	 5.1 	  & 	 44 	  & 	 10.0 	  & 	 19 	  \\
 22 	  & 	 4.4 	  & 	 153 	  & 	 4.9 	  & 	 45 	  & 	 3.1 	  & 	 15 	  & 	 4.9 	  & 	 56 	  & 	 4.3 	  & 	 37 	  \\
 37 	  & 	 4.6 	  & 	 153 	  & 	 5.1 	  & 	 45 	  & 	 3.1 	  & 	 15 	  & 	 4.6 	  & 	 56 	  & 	 4.4 	  & 	 37 	  \\
 90 	  & 	 4.9 	  & 	 72 	  & 	 4.6 	  & 	 18 	  & 	 3.7 	  & 	 4 	  & 	 5.0 	  & 	 34 	  & 	 8.1 	  & 	 16 	  \\
 230 	  & 	 5.1 	  & 	 33 	  & 	 3.4 	  & 	 9 	  & 	 1.2 	  & 	 1 	  & 	 5.2 	  & 	 16 	  & 	 7.0 	  & 	 7 	  \\
\hline 
\\
\hline
 $\nu$ [GHz]	  & \multicolumn{10}{c} {median relative peak flux density [Jy]} 	  \\
	  & 	 All 	  & 	 N 	  & 	 BLO 	  & 	 N 	  & 	 GAL  	  & 	 N 	  & 	 HPQ  	  & 	 N 	  & 	 LPQ 	  & 	 N 	  \\
\hline 
 4.8 	  & 	 1.8 	  & 	 85 	  & 	 1.7 	  & 	 34 	  & 	 1.1 	  & 	 8 	  & 	 1.8 	  & 	 33 	  & 	 3.8 	  & 	 10 	  \\
 8 	  & 	 2.4 	  & 	 99 	  & 	 2.1 	  & 	 38 	  & 	 1.3 	  & 	 8 	  & 	 2.2 	  & 	 41 	  & 	 5.3 	  & 	 12 	  \\
 14.5 	  & 	 2.7 	  & 	 117 	  & 	 2.4 	  & 	 45 	  & 	 1.6 	  & 	 9 	  & 	 3.1 	  & 	 44 	  & 	 4.6 	  & 	 19 	  \\
 22 	  & 	 2.6 	  & 	 151 	  & 	 2.8 	  & 	 44 	  & 	 1.8 	  & 	 15 	  & 	 2.8 	  & 	 56 	  & 	 2.5 	  & 	 36 	  \\
 37 	  & 	 2.8 	  & 	 151 	  & 	 3.1 	  & 	 44 	  & 	 2.0 	  & 	 15 	  & 	 2.8 	  & 	 56 	  & 	 2.5 	  & 	 36 	  \\
 90 	  & 	 3.1 	  & 	 72 	  & 	 2.8 	  & 	 18 	  & 	 2.4 	  & 	 4 	  & 	 3.6 	  & 	 34 	  & 	 4.8 	  & 	 16 	  \\
 230 	  & 	 3.6 	  & 	 31 	  & 	 1.9 	  & 	 9 	  & 	 $-$ 	  & 	 $-$ 	  & 	 3.6 	  & 	 15 	  & 	 6.2 	  & 	 7 	  \\
\hline 
\end{tabular}
\end{table*}

Considering the observed peak flux densities, there are no clear 
differences between the frequency bands according to the KW-analysis. When 
the relative peak flux densities are considered the 
KW-analysis shows that only the 4.8\,GHz differs from the 
other frequency bands. Also the 
higher frequency bands from 14.5\,GHz to 230\,GHz form one group while the 
8--22\,GHz frequency bands form another group. The median values have
an increasing trend in the relative peak flux densities as the frequency increases.
It also appears that in BLOs and HPQs the 
flux densities are somewhat higher at higher frequencies but the difference is not 
large. For LPQs, the peak flux densities at the lower 4.8--14.5\,GHz 
frequencies clearly differ from those at the higher frequency bands
but this is probably due to the much smaller number of sources at these 
frequencies. As was already noted in the previous section, at lower 
frequencies the flares are more superposed and it is the extreme objects that 
stand out. The fainter sources do not show distinguishable flares at 
lower frequencies and therefore are not included in the calculations.
This is also seen in the KW-analysis which shows that at 4.8 and 8\,GHz 
the LPQs differ from the other classes significantly. At 14.5\,GHz 
the difference with the higher frequencies is not as clear when the relative peak 
flux densities are considered. At the higher frequencies 
there are no significant differences between the source classes.

\subsection{Flare intensity and duration}
In some cases we could not calculate the duration 
of the flare because either the 
starting or ending point was not determined. This varied between the 
frequency bands so that we were able to 
calculate the duration for 150 flares at 37\,GHz and for 31 flares at 
230\,GHz. The durations of the flares varied from 0.3 years to 
13.2 years.

The duration seems to decrease with increasing frequency but again 
the difference is not very large, only 0.6 years at most between 4.8 and 90\,GHz.
According to the KW-analysis there are also no significant differences between the 
source classes even though the LPQs seem to have slightly longer durations.
This could indicate that there are differences in the physical conditions 
of the jets in LPQs and other classes. 
Of particular note is the long median duration of the flares: They last about 2.5 
years at 22 and 37\,GHz. This means that short multi-wavelength 
campaigns are not able to reveal the complete behaviour of these sources at radio 
frequencies. Long-term monitoring is needed to study the variability at these 
frequencies.

The rise and decay times appear to be approximately equal and so the 
flares are relatively symmetric. At 22 and 37\,GHz the median rise times 
are 1 year and decay 
times 1.3 years in accordance with the results of \cite{valtaoja99}. 
There are no significant differences between the frequency bands 
but the decay times of the LPQs seem to be slightly longer which is 
also seen in the KW-analysis.

\begin{figure}
\resizebox{\hsize}{!}{\includegraphics{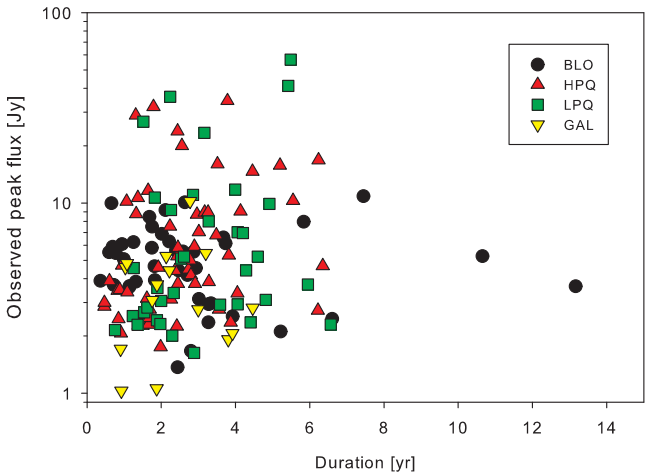}}
\caption{Observed peak flux density against the duration of the flare at 37\,GHz.}
\label{peak_37}
\end{figure}

In Fig. \ref{peak_37} we plotted the observed peak flux density of the 
flares against the duration for all sources at 37\,GHz. 
The Spearman rank correlation 
for these parameters is $r=0.15$ with probability $p=0.03$. Thus the 
correlation is not strong. At lower frequencies the correlation was slightly
higher, decreasing towards higher frequencies. If only quasars (HPQs and LPQs) 
are considered, there is a correlation of $r=0.29$ with $p=0.002$ at 37\,GHz.
The result is similar if HPQs and LPQs are treated separately.
For BLOs there are no significant correlations \citep{nieppola08}.

There is a slightly higher positive correlation if the relative peak flux density 
is studied 
against the duration but the distribution of points looks very similar. 
At 37\,GHz $r=0.26$ 
with $p=0.001$. Again the correlations at lower frequencies are slightly 
higher, decreasing towards higher frequencies. This time the correlation 
of quasars is even higher with $r=0.36$ and $p=0.0002$. Also in this case 
there are no correlations between these parameters if only BLOs are 
considered \citep{nieppola08}.

There is a selection effect which could affect the results. There are fewer 
weak flares of long duration. If such a 
flare is seen on the flux density curve, it may get overlooked as it resembles 
more a secular baseline flux density trend than a flare. Also the 
shorter the flares 
are, the more they show up like spikes in the flux density curve and are 
taken as flares more easily. 

\begin{figure}
\resizebox{\hsize}{!}{\includegraphics{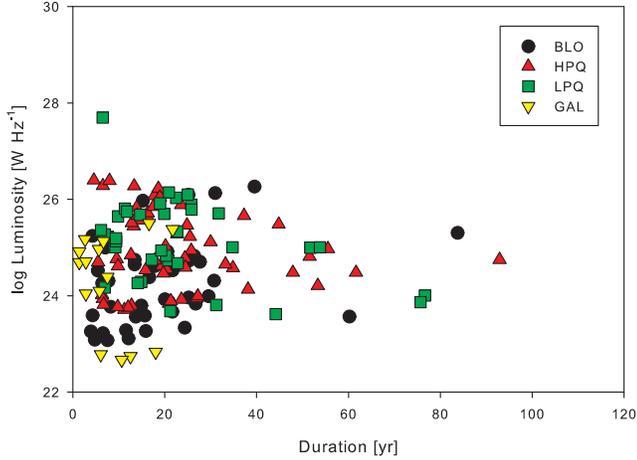}}
\caption{Corrected peak luminosity at 37\,GHz against the duration of the flare corrected for redshift and Doppler boosting.}
\label{corr_dur_37}
\end{figure}

One other significant effect is Doppler boosting. When the sources are 
more boosted, their flux density levels are higher and at the same time the 
flares appear shorter. Therefore we corrected the peak flux density and duration 
for redshift and Doppler boosting using Eqs. \ref{eq:dur} and \ref{eq:lum} 
\citep[e.g.][]{kembhavi99}

\begin{equation} \label{eq:dur}
 t_i = t_0\frac{D_\mathrm{var}}{1+z}
\end{equation}

\begin{equation} \label{eq:lum}
 L_i = S_\mathrm{0}\frac{4\pi D_\mathrm{L}}{1+z}\left(\frac{1+z}{D_\mathrm{var}}\right)^3,
\end{equation}
where $t$ is the duration,  $S_\mathrm{0}$ is the flux density, $D_\mathrm{L}$ 
is the luminosity distance, $z$ is the redshift and 
$D_\mathrm{var}$ is the Doppler boosting factor taken from Hovatta et al. 
(in preparation) where also other jet parameters such as jet speed are 
studied in more detail.
When the corrected durations are studied there are still no clear differences 
between the frequency bands according to the KW-analysis. The LPQs have slightly longer 
flares but the difference is smaller than in the uncorrected case. Instead, at the 
lower 4.8--14.5\,GHz frequencies the BLOs differ from the other source classes due to their shorter flares.
The corrected peak luminosity against the corrected duration is shown in 
Fig. \ref{corr_dur_37}. The positive correlation disappears and there 
are no significant correlations according to the Spearman rank correlation.

\subsection{Correspondence to the generalised shock model}
According to the generalised shock model \citep{valtaoja92}, the observed 
properties of a flare depend on the observing frequency and whether the peak 
frequency of the flare is higher or lower than the observing frequency. We 
studied how well these 159 flares correspond to the model by calculating 
a few parameters for each flare. We point out, however, that in this work 
we have not extracted separate shock events from flares but treated each 
longer outburst event as an individual flare.

\begin{figure}
\resizebox{\hsize}{!}{\includegraphics{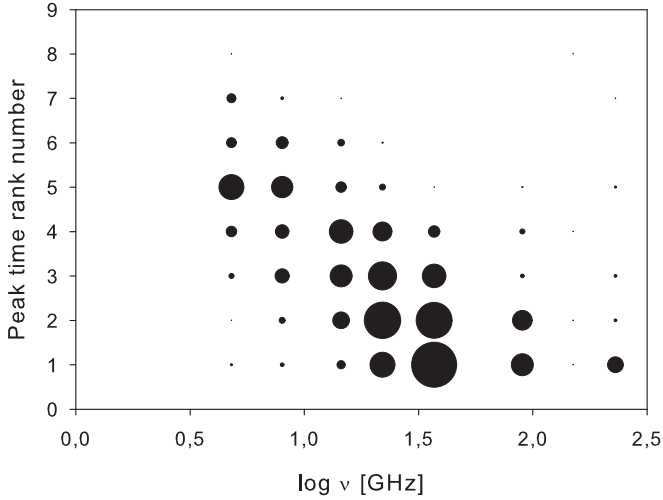}}
\caption{Peak time rank number plotted against the frequency. The frequency band having the flare peak first has been ranked 1.}
\label{bubble_time}
\end{figure}

First we determined the frequency band in which the flare occurs first and we 
have given 
it rank number 1. The second frequency band then gets number 2 and so on. 
For example, if the flare peak is first at 90\,GHz, then at 22\,GHz and then at 
37\,GHz, the rank numbers are 1 for 90\,GHz, 2 for 22\,GHz and 3 for 37\,GHz.
In this procedure we ignored flares which are clearly multi-peaked and for 
which it is impossible to determine the actual peak at all frequency bands. 
We have not 
ignored frequencies with larger gaps and therefore we expect scatter in the 
results. We plotted the peak time rank numbers against the frequency 
in Fig. \ref{bubble_time} in a bubble plot. The bigger the bubble, the 
more cases there are in that rank number and frequency. The plot shows that, 
as expected, the flares are peaking first at higher frequencies 
but there are also outliers and scatter.
 
\begin{figure}
\resizebox{\hsize}{!}{\includegraphics{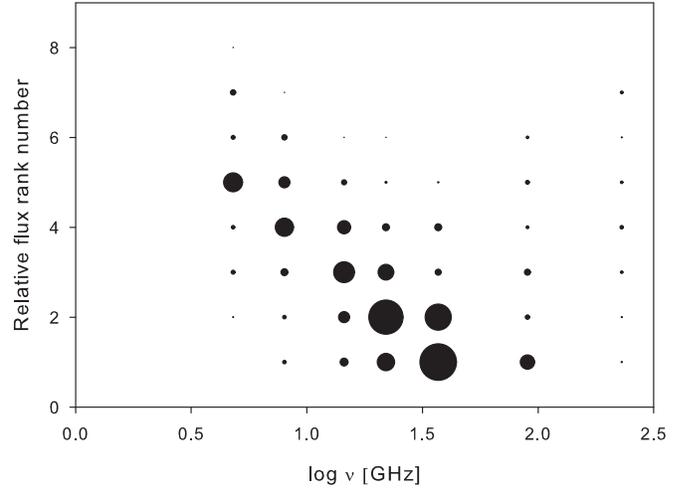}}
\caption{Relative flux density rank number plotted against the frequency. Frequency with highest relative flux density has been ranked 1.}
\label{bubble_rel}
\end{figure}

Another prediction of the model is that relative flux density should be highest at the 
peak frequency, which in the case of high-peaking flares should be in the upper 
part of our frequency range. We made a similar bubble plot of the 
relative flux density rank number against the frequency (Fig. \ref{bubble_rel}).
In this case we have given rank number 1 to the frequency at which the 
relative flux density is highest. Rank number 2 is given to the frequency which has 
the second highest relative flux density and so on. Fig. \ref{bubble_rel} shows that 
the highest flux density is usually at 22 and 37\,GHz, although the 
number of flares at higher frequencies is lower and therefore the result 
could be slightly biased. Also the sampling of the higher frequencies is more 
sparse and therefore the actual flare peaks could have been missed in the 
observations. 

A more straightforward way of studying the correspondence between the model 
and the observations is to normalise the relative flux density to the flux density of the 
peak frequency $\nu_\mathrm{max}$ of the flare. 
In our analysis we have simply taken $\nu_\mathrm{max}$ as the 
frequency band at which the relative peak flux density is highest. A more 
accurate way would be to interpolate the peak frequency but we find it 
unsuitable for our data which has large gaps especially at the higher 
90 and 230\,GHz frequency bands, making the estimation of the true peak time 
and peak flux density of the flare difficult.
According to \cite{valtaoja92}
the maximum flux density of the flare should correspond to the plateau stage in the 
generalised model. Frequencies higher than $\nu_\mathrm{max}$ correspond to 
the growth stage and should have 
$\Delta S_\mathrm{max} \propto \nu^{\alpha_{\mathrm{thin}}}$. 
Frequencies lower than $\nu_\mathrm{max}$ should have 
$\Delta S_\mathrm{max} \propto \nu^{\mathrm{b}}$, 
where $\mathrm{b}$ depends on the model. This approach was originally used by 
\cite{valtaoja92} and 
later by \cite{lainela94}. They used a smaller sample of sources and as 
the sources had been monitored for only a short time, they used the overall 
minimum and maximum values ever observed as the relative flux density. Our approach 
should be more accurate as the flares were individually separated and 
the number of flares is much higher.

\begin{figure}
\resizebox{\hsize}{!}{\includegraphics{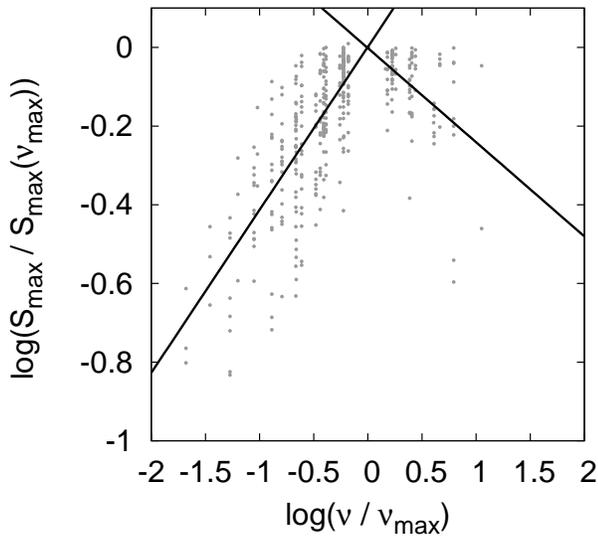}}
\caption{Observed maximum relative flux density scaled to the frequency at which the maximum occurs. The two straight lines give best fits to the section below and above the peak frequency.}
\label{shockmodel}
\end{figure}

In Fig. \ref{shockmodel} we plotted the 
normalised relative flux density against the normalised frequency. All the values 
were normalised to the frequency at which the relative flux density is the highest.
We also fitted straight lines to below and above the peak frequency.
We can see that there is large scatter in the values but the trend is very 
much as expected when compared to the model in \cite{valtaoja92}. 
The fit gives $\Delta S_\mathrm{max} \propto \nu^{0.41}$
for the rising part and $\Delta S_\mathrm{max} \propto \nu^{-0.24}$ for the 
declining part. We also fitted the data separately for quasars and HPQs and LPQs 
and the results were very similar to those of the whole sample. According to the 
model, there should be a plateau between the slopes which is not seen in our 
analysis. We also tried to fit a plateau stage into the model, but two 
linear components describe the data more accurately.
The same result was already obtained in \cite{valtaoja92}. They got 
a slope of 0.52 for the rising part and $-0.26$ for the declining part, which are 
very close to the values obtained here. \cite{stevens94} also obtained a value of
$-0.28$ for the declining part in their study of 17 blazars between 22 and 375\,GHz.

Time delays between the frequencies are also important but more difficult to 
study. This is because the sampling is often sparse and the actual peak time
could have been missed during observations. If a source is observed once in 
two weeks, which is quite a common sampling rate for these sources at 
Mets\"ahovi, the peak can occur at any time within a month. 
Often the gaps are longer due to weather constrains or maintenance. 
Therefore any time delays shorter than a month are not reliable.
At the higher 90 and 230\,GHz frequencies the situation is even worse because the 
flares are faster and the data acquisition is even more sparse.

\begin{figure}
\resizebox{\hsize}{!}{\includegraphics{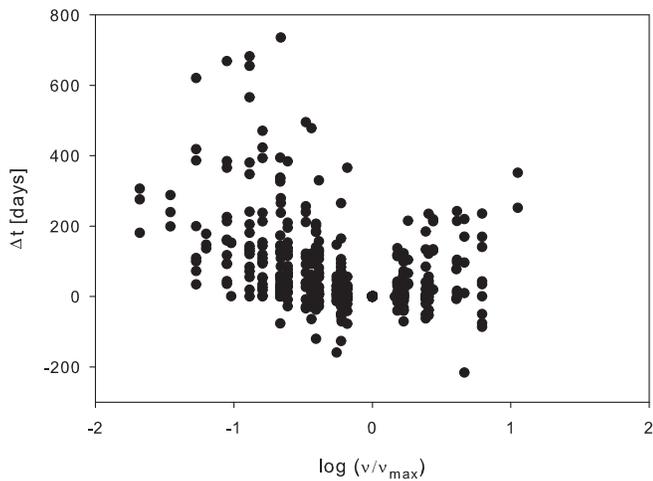}}
\caption{Time delay between the observing frequency and peak frequency against the normalised frequency.}
\label{timedelays}
\end{figure}

We calculated median time delays at different frequency bands 
and for 37\,GHz 
it is 34 days, increasing towards lower frequencies and being 181 days at 4.8\,GHz. 
According to the generalised shock model, there should not be significant 
time delays above the peak frequency but below it they should increase. 
We calculated the time delays between the peak frequency and 
other frequencies and 
plotted them against the normalised frequency in Fig. \ref{timedelays}. 
It is clear that the trend is correct with time delays being larger at 
frequencies lower than the peak frequency. There is also large scatter in 
the values which is expected because the time delays are so difficult to 
determine especially at higher frequencies. In this analysis we took into 
account all flares where the peak was not clearly different at different 
frequencies and we did not ignore flares with poor sampling. Even time 
delays as large as 300 days can be due to poor sampling at 230\,GHz. However, 
if compared to the model in \cite{valtaoja92}, we see clear correspondence 
between the model and observations. 

\begin{figure}
\resizebox{\hsize}{!}{\includegraphics{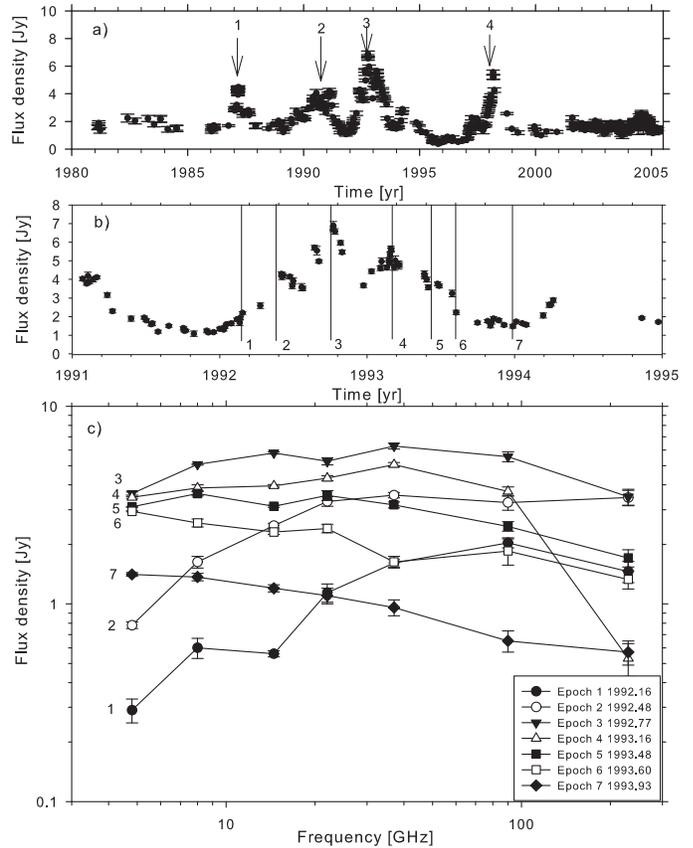}}
\caption{a) Flux density curve of the BLO source \object{0235+164} at 37\,GHz with each flare analysed marked with arrows. b) Flare number 3 in more detail. c) Spectra of every epoch marked in Fig b).}
\label{0235}
\end{figure}

We also studied one individual source, \object{0235+164}, and its flare in 
1993 in more detail. Figure \ref{0235}a, shows the flux density curve of the source at 
37\,GHz, with every flare analysed for this source marked using arrows. Fig. 
\ref{0235}b shows the 1993 flare in more detail at 37\,GHz. We have marked seven 
epochs of the flare and constructed simultaneous spectra at every epoch at all 
the frequencies used in the analysis (Fig. \ref{0235}c). The spectra are 
simultaneous within two weeks 
but the 230\,GHz points are in general more uncertain because of the more sparse 
sampling. The flare follows the shock model quite well with the flux density first rising 
at all frequencies. When the flux density starts to decay the peak of the spectrum moves
to lower frequencies. No clear plateau is seen in this flare which agrees with the 
general results for the whole sample.

\section{Discussion}\label{sec:discussion}

We have studied the observational properties of radio flares for
a large sample of densely monitored variable AGNs. For many of
the sources the data sets span more than 25 years. For this
study we only chose objects with clearly distinguishable
radio flares, which biases this sample towards sources
with notable variability and flares with relatively simple
morphology.

Within our sample, the BLOs have higher variability indices than quasars. This
is in part caused by the fact that only
the BLOs with truly distinct variability features
were included in this study. Compared to the variability
indices obtained in \cite{nieppola07}, our indices are higher,
the median variability indices of BLOs being 0.78 in this study
and 0.33 in \cite{nieppola07}.

The sample of \cite{nieppola07} also includes fainter BLOs for which no
well-structured radio flares could be extracted from the flux density
curves. The differences between the variability behaviour of these two
BLO samples can be explained by the fact that only a small minority of
the full BLO sample actually shows any extreme behaviour in the radio
domain, most of them being faint and relatively non-variable even over
several years of monitoring observations.  However, it is noteworthy
that the large BLO sample of \cite{nieppola07} only includes 5 years of
data compared to our 25 years, and at least some of the sources may
radically change their behaviour over longer timescales, as discussed
in \cite{hovatta07}.

Earlier studies of higher 90 and 230\,GHz variability indices were 
done by \cite{tornikoski93} and \ \cite{tornikoski00}. Their sample 
included southern AGNs observed with the SEST telescope. The variability 
indices are significantly lower than what we have obtained (for 
comparison we calculated the variability index defined as 
$(S_\mathrm{max}-S_\mathrm{min})/S_\mathrm{min}$ for our sample) but also 
the number of datapoints is much smaller. The average number of 
datapoints in our sample is almost five times larger. In \cite{tornikoski00}
the effect of the number of datapoints was studied and they noticed that 
the variability indices increased with increasing number of datapoints, 
confirming our findings where only very long-term monitoring reveals the 
true minimum and maximum flux densities of most sources. Even though the highest 
flux densities during flares are typically observed in the 
37 and 22\,GHz frequency bands (as seen in Fig. \ref{bubble_rel}), the 
90\,GHz variability indices obtained in this paper are of the same order as 
those in the lower frequency bands, indicating that strong flares can 
indeed occur also in the mm-domain.

When we studied the duration and intensity of the individual
flares, it was evident that most flares are very long lasting.
The median duration of flares at 22 and 37\,GHz is 2.5 years,
which means that short-term observing campaigns are rarely helpful for
describing the ``typical behaviour" of these sources.
When we add to this the result from \cite{hovatta07}
that for these sources flares occur on average every 4--6 years,
we can conclude that monitoring campaigns lasting at least
ca. 5--7 years are needed in order to catch a typical
source both in a quiescent and flaring state and to follow
the flare evolution from the beginning to the end.

The flare intensities increase with flare duration only to
some degree. For flares lasting 2--3 years we can see almost
as high peak flux densities (observed peak flux density as well as the
relative flux density during the flare) as for those lasting a few years longer.
A striking result is that when we calculate the Doppler-corrected
peak luminosities of these flares, the weak positive correlation
disappears altogether and there is no correlation between
the Doppler corrected peak luminosity against the Doppler corrected flare
duration. This means that the energy
release in a flare does not increase with duration for these flares.

When studying the rise and fall times of the flares we see
that flares in general have semi-symmetric shapes,
with the decay time typically being 1.3 times the rise time.
The morphologies of flares are typically not very simple, though,
and flares with multiple peaks or other finer structure
complicate the analysis. In general, flares at 37\,GHz or lower frequencies
rarely consist of smooth rise and decay components only, and a more thorough
analysis is needed to fully understand
the physical characteristics of radio flares and to
separate the individual flare components (Hovatta et al.
in preparation).

Based on the observational results of this study, we have studied 
the correspondence between the observations and the generalised 
shock model by \cite{valtaoja92}. For 99 flares we had an estimate of the time 
delays between at least 3 frequency bands. Nearly half of them (46 flares) 
seemed to behave as the shock model predicts with no time delays above the 
peak frequency and increasing time delays below it. In addition in 34 cases 
the time delays were shorter than the sampling rate and therefore we 
cannot say whether they follow the shock model or not. Only in 20\% (19 cases) 
the time delays and amplitudes did not follow the shock model. In these cases, 
however, the flares were defined to have multiple peaks and therefore the 
time delays were difficult to determine. Also the sampling rate 
may have been poor. In any case the flares seem to follow the shock model 
in general, even though there is large scatter due to poor sampling and 
complicated structure of the flares at different frequency bands. Of the 
well-determined flares where we could study if they are high- or low-peaking, 
25 were high-peaking and 3 low-peaking at 22 and 37\,GHz.
This indicates that the peak frequency of the flares in these sources is 
somewhere in the millimetre domain. This is in agreement with \cite{stevens94} who 
found that in their sample of 17 blazars studied at 22--375\,GHz the peak was 
at 90\,GHz or below.
Our results for the whole sample are 
very similar to those for BLOs studied in \cite{nieppola08}.

When we studied the shape of the flare spectra normalised to the peak 
frequency, we found the best fit to be  
$\Delta S_\mathrm{max} \propto \nu^{0.41}$ for the rising part and 
$\Delta S_\mathrm{max} \propto \nu^{-0.24}$ for the declining part. 
We did not find a 
plateau between the slopes which is contradictory to the model but in 
agreement with earlier studies by \cite{valtaoja92} and \cite{lindfors06}. 
The declining slope 
$\alpha = -0.24$ corresponds to the optically thin spectral index 
$\alpha_{\mathrm{thin}}$ for a power-law electron energy distribution 
$N(E) \propto N_0E^{-s}$ with $s = 1.5$ \citep{valtaoja92}. 
This is 
slightly flatter than what is obtained for \object{3C 273} in 
\cite{turler00} and for \object{3C 279} in \cite{lindfors06}, where values 
close to $s = 2$ were 
obtained. In their studies the flare spectra were flattening from 
$\alpha_{\mathrm{thin}} = -1.1$ to $\alpha_{\mathrm{thin}} = -0.5$ as the 
flare evolved. This is in accordance with the original shock model by 
\cite{marscher85}. In \cite{stevens94} the declining slope of $-0.28$ was explained 
as being due to the radiation being only partially thin close to the 
turnover frequency. This would cause the points to deviate from a straight line 
in the declining part of the spectrum. Also, 
data at sub-mm frequencies above 230\,GHz would be required to estimate 
the accurate value of $\alpha_{\mathrm{thin}}$. The rising slope of 0.41 is 
in agreement with the original shock model \citep{marscher85} and also independently 
obtained in 
\cite{valtaoja92}. Therefore we consider 
our results to be in good agreement with the models, considering the scatter in our data.

\section{Conclusions}\label{sec:conclusions}
In this paper we studied the long-term radio variability of a sample of 
55 AGNs. We divided the flux density curves into individual flares and 
studied the properties of 159 well-monitored flares in at least two 
frequency bands between 4.8 and 230\,GHz. Our main results can be 
summarised as follows:

\begin{enumerate}
\item The flares last on average 2.5 years at 22 and 37\,GHz and are 
only slightly longer at lower frequencies. This has important consequences when 
planning multifrequency campaigns, which usually last only from a couple of days to 
months and therefore are unable to capture a radio flare as a whole.

\item There are no significant differences between the different source classes when 
the durations of the flares are studied although there is weak evidence for 
longer flares in the LPQs. When Doppler-corrected durations are studied the 
LPQs still have the longest flares but the difference is smaller. 
Also, the BLOs with their shorter flares differ from other source classes 
at the lower 4.8--14.5\,GHz frequencies.

\item When studying the duration of the flares against the intrinsic 
Doppler-corrected peak luminosity, we found no correlation indicating that the energy release in a flare does not increase with duration for these flares.

\item The flares seem to follow the generalised shock model by 
\cite{valtaoja92} quite well even though there is large scatter due to 
incomplete sampling and complicated structure of the flares at different 
frequency bands.
\end{enumerate}

\begin{acknowledgements}
We acknowledge the support of the Academy of Finland (project numbers 212656 and 
210338). UMRAO is supported in part by a series of grants from the NSF, most recently AST-0607523, and by funds 
from the University of Michigan Department of Astronomy.
\end{acknowledgements}

\bibliographystyle{/home/tho/texmf/tex/aa-package/bibtex/aa}
\bibliography{/home/tho/texmf/tex/aa-package/bibtex/thbib}

\longtabL{1}{
\begin{landscape}
\begin{longtable}{lllcllllllllllll}
\caption{\label{table:sourcelist}The sample. For each source the number of flares, the number of datapoints since 1980 in each frequency band, the maximum flux density ($S_{max}$), variability index V and median duration of flares at 37\,GHz are listed. Colum 4 indicates if the source is a high-confidence EGRET detection (x) or a possible detection (?).}\\
\hline
\hline
B1950 	  & 	 Other 	  & 	 Class 	  & EGRET	 &     Number 	  & 	 \multicolumn{8}{c}{Number of datapoints since 1980 in each frequency band}  	  & 	 $S_{max}$ [Jy]	  & 	 V 	  & 	 Med duration [yr]	  \\
 name 	  & 	 name 	  & 	  	  & detection	 &   of flares 	  & 	 4.8 	  & 	 8 	  & 	 14.5 	  & 	 22 	  & 	 37 	  & 	 90 	  & 	 150 	  & 	 230 	  & 	 37\,GHz 	  & 	 37\,GHz 	  & 	 37\,GHz 	  \\
\hline
\endfirsthead
\caption{Continued.}\\
\hline
\hline
B1950 	  & 	 Other 	  & 	 Class 	  & EGRET	 &	 Number 	  & 	 \multicolumn{8}{c}{Number of datapoints since 1980 in each frequency band}  	  & 	 $S_{max}$ [Jy]	  & 	 V 	  & 	 Med duration [yr]	  \\
 name 	  & 	 name 	  & 	  	  & detection	 &	 of flares 	  & 	 4.8 	  & 	 8 	  & 	 14.5 	  & 	 22 	  & 	 37 	  & 	 90 	  & 	 150 	  & 	 230 	  & 	 37\,GHz 	  & 	 37\,GHz 	  & 	 37\,GHz 	  \\
\hline 
\endhead
\hline
\endfoot
 \object{0007$+$106} 	  & 	 III ZW 2 	  & 	 GAL 	  & 	  	  & 	 6 	  & 	 234 	  & 	 438 	  & 	 477 	  & 	 309 	  & 	 253 	  & 	 26 	  & 	 - 	  & 	 15 	  & 	 3.08 	  & 	 0.84 	  & 	 1.82 	  \\
 \object{0059$+$581} 	  & 	  	  & 	 LPQ 	  & 	  	  & 	 2 	  & 	 - 	  & 	 - 	  & 	 - 	  & 	 148 	  & 	 101 	  & 	 - 	  & 	 - 	  & 	 - 	  & 	 5.23 	  & 	 0.70 	  & 	 2.47 	  \\
 \object{0106$+$013} 	  & 	 OC 012 	  & 	 HPQ 	  & 	  	  & 	 3 	  & 	 106 	  & 	 256 	  & 	 220 	  & 	 197 	  & 	 160 	  & 	 - 	  & 	 - 	  & 	 - 	  & 	 3.77 	  & 	 0.68 	  & 	 3.87 	  \\
 \object{0109$+$224} 	  & 	 S2 0109+22 	  & 	 BLO 	  & 	  	  & 	 3 	  & 	 114 	  & 	 152 	  & 	 162 	  & 	 181 	  & 	 152 	  & 	 - 	  & 	 - 	  & 	 - 	  & 	 3.13 	  & 	 0.78 	  & 	 3.01 	  \\
 \object{0208$-$512} 	  & 	  	  & 	 BLO 	  & 	 x 	  & 	 1 	  & 	 - 	  & 	 - 	  & 	 - 	  & 	 - 	  & 	 - 	  & 	 57 	  & 	 - 	  & 	 31 	  & 	 $6.51^1$ 	  & 	 $0.66^1$ 	  & 	 $6.52^1$ 	  \\
 \object{0224$+$671} 	  & 	  	  & 	 LPQ 	  & 	  	  & 	 2 	  & 	 110 	  & 	 115 	  & 	 114 	  & 	 106 	  & 	 80 	  & 	 42 	  & 	 - 	  & 	 - 	  & 	 3.06 	  & 	 0.65 	  & 	 1.99 	  \\
 \object{0234$+$285} 	  & 	 4C2 8.07 	  & 	 HPQ 	  & 	 x 	  & 	 2 	  & 	 221 	  & 	 228 	  & 	 264 	  & 	 145 	  & 	 105 	  & 	 80 	  & 	 - 	  & 	 - 	  & 	 4.68 	  & 	 0.48 	  & 	 4.57 	  \\
 \object{0235$+$164} 	  & 	  	  & 	 BLO 	  & 	 x 	  & 	 4 	  & 	 723 	  & 	 1024 	  & 	 946 	  & 	 399 	  & 	 548 	  & 	 160 	  & 	 26 	  & 	 46 	  & 	 6.88 	  & 	 0.88 	  & 	 2.52 	  \\
 \object{0333$+$321} 	  & 	 NRAO 140 	  & 	 LPQ 	  & 	  	  & 	 2 	  & 	 - 	  & 	 - 	  & 	 - 	  & 	 178 	  & 	 153 	  & 	 - 	  & 	 - 	  & 	 - 	  & 	 2.68 	  & 	 0.56 	  & 	 5.49 	  \\
 \object{0336$-$019} 	  & 	 CTA 026 	  & 	 HPQ 	  & 	 x 	  & 	 2 	  & 	 178 	  & 	 506 	  & 	 433 	  & 	 108 	  & 	 82 	  & 	 67 	  & 	 - 	  & 	 - 	  & 	 3.86 	  & 	 0.51 	  & 	 2.45 	  \\
 \object{0415$+$379} 	  & 	 3C 111 	  & 	 GAL 	  & 	 ? 	  & 	 3 	  & 	 232 	  & 	 471 	  & 	 425 	  & 	 147 	  & 	 88 	  & 	 29 	  & 	 - 	  & 	 - 	  & 	 10.22 	  & 	 0.65 	  & 	 2.13 	  \\
 \object{0420$-$014} 	  & 	 OA 129 	  & 	 HPQ 	  & 	 x 	  & 	 3 	  & 	 614 	  & 	 1042 	  & 	 917 	  & 	 378 	  & 	 417 	  & 	 198 	  & 	 - 	  & 	 59 	  & 	 15.68 	  & 	 0.76 	  & 	 3.48 	  \\
 \object{0422$+$004} 	  & 	 OF 038 	  & 	 BLO 	  & 	  	  & 	 2 	  & 	 165 	  & 	 516 	  & 	 373 	  & 	 144 	  & 	 131 	  & 	 44 	  & 	 - 	  & 	 20 	  & 	 2.40 	  & 	 0.76 	  & 	 1.96 	  \\
 \object{0430$+$052} 	  & 	 3C 120 	  & 	 GAL 	  & 	  	  & 	 4 	  & 	 621 	  & 	 1102 	  & 	 1003 	  & 	 417 	  & 	 462 	  & 	 109 	  & 	 - 	  & 	 - 	  & 	 5.45 	  & 	 0.54 	  & 	 2.05 	  \\
 \object{0446$+$112} 	  & 	 PKS 0446+112 	  & 	 GAL 	  & 	 ? 	  & 	 2 	  & 	 - 	  & 	 - 	  & 	 - 	  & 	 119 	  & 	 70 	  & 	 - 	  & 	 - 	  & 	 - 	  & 	 2.76 	  & 	 0.73 	  & 	 3.46 	  \\
 \object{0528$+$134} 	  & 	 PKS 0528+134 	  & 	 LPQ 	  & 	 x 	  & 	 4 	  & 	 215 	  & 	 500 	  & 	 426 	  & 	 548 	  & 	 371 	  & 	 130 	  & 	 - 	  & 	 - 	  & 	 11.03 	  & 	 0.72 	  & 	 2.70 	  \\
 \object{0537$-$441} 	  & 	  	  & 	 HPQ 	  & 	 x 	  & 	 1 	  & 	 - 	  & 	 - 	  & 	 - 	  & 	 - 	  & 	 - 	  & 	 50 	  & 	 - 	  & 	 25 	  & 	 $7.99^1$ 	  & 	 $0.64^1$ 	  & 	 - 	  \\
 \object{0607$-$157} 	  & 	  	  & 	 LPQ 	  & 	  	  & 	 1 	  & 	 731 	  & 	 951 	  & 	 1002 	  & 	 - 	  & 	 - 	  & 	 75 	  & 	 - 	  & 	 27 	  & 	 $9.65^1$ 	  & 	 $0.62^1$ 	  & 	 $1.75^1$ 	  \\
 \object{0642$+$449} 	  & 	 OH 471 	  & 	 LPQ 	  & 	  	  & 	 1 	  & 	 - 	  & 	 - 	  & 	 - 	  & 	 270 	  & 	 219 	  & 	 47 	  & 	 - 	  & 	 - 	  & 	 3.94 	  & 	 0.44 	  & 	 3.59 	  \\
 \object{0716$+$714} 	  & 	  	  & 	 BLO 	  & 	 x 	  & 	 2 	  & 	 369 	  & 	 347 	  & 	 689 	  & 	 203 	  & 	 493 	  & 	 - 	  & 	 - 	  & 	 - 	  & 	 6.28 	  & 	 0.90 	  & 	 3.07 	  \\
 \object{0735$+$178} 	  & 	 PKS 0735+17 	  & 	 BLO 	  & 	 x 	  & 	 1 	  & 	 416 	  & 	 640 	  & 	 693 	  & 	 309 	  & 	 295 	  & 	 99 	  & 	 - 	  & 	 38 	  & 	 5.26 	  & 	 0.80 	  & 	 10.66 	  \\
 \object{0736$+$017} 	  & 	  	  & 	 HPQ 	  & 	  	  & 	 2 	  & 	 137 	  & 	 320 	  & 	 294 	  & 	 208 	  & 	 157 	  & 	 79 	  & 	 - 	  & 	 26 	  & 	 4.71 	  & 	 0.75 	  & 	 1.30 	  \\
 \object{0754$+$100} 	  & 	 OI 090.4 	  & 	 BLO 	  & 	  	  & 	 2 	  & 	 122 	  & 	 295 	  & 	 289 	  & 	 208 	  & 	 169 	  & 	 24 	  & 	 - 	  & 	 - 	  & 	 2.94 	  & 	 0.67 	  & 	 3.27 	  \\
 \object{0804$+$499} 	  & 	  	  & 	 HPQ 	  & 	 ? 	  & 	 4 	  & 	 206 	  & 	 209 	  & 	 199 	  & 	 257 	  & 	 160 	  & 	 25 	  & 	 - 	  & 	 - 	  & 	 3.47 	  & 	 0.79 	  & 	 0.83 	  \\
 \object{0827$+$243} 	  & 	 OJ 248 	  & 	 LPQ 	  & 	 ? 	  & 	 3 	  & 	 - 	  & 	 - 	  & 	 - 	  & 	 133 	  & 	 62 	  & 	 16 	  & 	 - 	  & 	 - 	  & 	 2.93 	  & 	 0.62 	  & 	 1.36 	  \\
 \object{0847$-$120} 	  & 	 J0850-1213 	  & 	 LPQ 	  & 	 x 	  & 	 3 	  & 	 - 	  & 	 - 	  & 	 - 	  & 	 331 	  & 	 131 	  & 	 - 	  & 	 - 	  & 	 - 	  & 	 3.66 	  & 	 0.69 	  & 	 1.54 	  \\
 \object{0851$+$202} 	  & 	 OJ 287 	  & 	 BLO 	  & 	 x 	  & 	 9 	  & 	 762 	  & 	 902 	  & 	 955 	  & 	 895 	  & 	 916 	  & 	 272 	  & 	 - 	  & 	 74 	  & 	 9.18 	  & 	 0.74 	  & 	 1.31 	  \\
 \object{1055$+$018} 	  & 	 OL 093 	  & 	 HPQ 	  & 	  	  & 	 4 	  & 	 354 	  & 	 573 	  & 	 639 	  & 	 255 	  & 	 229 	  & 	 80 	  & 	 - 	  & 	 27 	  & 	 7.05 	  & 	 0.60 	  & 	 2.61 	  \\
 \object{1156$+$295} 	  & 	 4C 29.45 	  & 	 HPQ 	  & 	 x 	  & 	 6 	  & 	 414 	  & 	 810 	  & 	 667 	  & 	 404 	  & 	 326 	  & 	 73 	  & 	 - 	  & 	 - 	  & 	 5.32 	  & 	 0.79 	  & 	 1.59 	  \\
 \object{1222$+$216} 	  & 	 PKS 1222+216 	  & 	 LPQ 	  & 	 x 	  & 	 2 	  & 	 - 	  & 	 - 	  & 	 - 	  & 	 243 	  & 	 116 	  & 	 27 	  & 	 - 	  & 	 14 	  & 	 2.65 	  & 	 0.74 	  & 	 2.38 	  \\
 \object{1226$+$023} 	  & 	 3C 273 	  & 	 LPQ 	  & 	 x 	  & 	 5 	  & 	 720 	  & 	 1033 	  & 	 951 	  & 	 939 	  & 	 1041 	  & 	 352 	  & 	 28 	  & 	 166 	  & 	 56.72 	  & 	 0.68 	  & 	 3.17 	  \\
 \object{1253$-$055} 	  & 	 3C 279 	  & 	 HPQ 	  & 	 x 	  & 	 6 	  & 	 774 	  & 	 1065 	  & 	 1039 	  & 	 762 	  & 	 790 	  & 	 234 	  & 	 28 	  & 	 122 	  & 	 34.48 	  & 	 0.66 	  & 	 2.50 	  \\
 \object{1308$+$326} 	  & 	 AU CV n 	  & 	 BLO 	  & 	  	  & 	 2 	  & 	 609 	  & 	 948 	  & 	 779 	  & 	 378 	  & 	 315 	  & 	 - 	  & 	 - 	  & 	 - 	  & 	 4.16 	  & 	 0.77 	  & 	 13.17 	  \\
 \object{1334$-$127} 	  & 	  	  & 	 HPQ 	  & 	 x 	  & 	 1 	  & 	 482 	  & 	 821 	  & 	 649 	  & 	 - 	  & 	 - 	  & 	 116 	  & 	 - 	  & 	 35 	  & 	 $9.35^1$ 	  & 	 $0.60^1$ 	  & 	 $2.41^1$ 	  \\
 \object{1413$+$135} 	  & 	  	  & 	 BLO 	  & 	  	  & 	 2 	  & 	 279 	  & 	 701 	  & 	 562 	  & 	 243 	  & 	 186 	  & 	 71 	  & 	 - 	  & 	 - 	  & 	 4.55 	  & 	 0.85 	  & 	 4.77 	  \\
 \object{1502$+$106} 	  & 	 OR 103 	  & 	 HPQ 	  & 	  	  & 	 3 	  & 	 - 	  & 	 - 	  & 	 - 	  & 	 156 	  & 	 136 	  & 	 - 	  & 	 - 	  & 	 - 	  & 	 2.28 	  & 	 0.57 	  & 	 1.98 	  \\
 \object{1510$-$089} 	  & 	 PKS 1510-089 	  & 	 HPQ 	  & 	 x 	  & 	 6 	  & 	 587 	  & 	 984 	  & 	 1081 	  & 	 245 	  & 	 263 	  & 	 111 	  & 	 - 	  & 	 39 	  & 	 5.83 	  & 	 0.74 	  & 	 1.50 	  \\
 \object{1606$+$106} 	  & 	 4C 10.45 	  & 	 LPQ 	  & 	 x 	  & 	 2 	  & 	 - 	  & 	 - 	  & 	 - 	  & 	 150 	  & 	 108 	  & 	 - 	  & 	 - 	  & 	 - 	  & 	 3.73 	  & 	 0.70 	  & 	 4.12 	  \\
 \object{1611$+$343} 	  & 	 DA 406 	  & 	 LPQ 	  & 	 x 	  & 	 1 	  & 	 - 	  & 	 - 	  & 	 250 	  & 	 229 	  & 	 198 	  & 	 - 	  & 	 - 	  & 	 - 	  & 	 5.23 	  & 	 0.54 	  & 	 4.60 	  \\
 \object{1633$+$382} 	  & 	 4C 38.41 	  & 	 LPQ 	  & 	 x 	  & 	 2 	  & 	 - 	  & 	 - 	  & 	 453 	  & 	 457 	  & 	 466 	  & 	 - 	  & 	 - 	  & 	 - 	  & 	 7.69 	  & 	 0.71 	  & 	 3.05 	  \\
 \object{1641$+$399} 	  & 	 3C 345 	  & 	 HPQ 	  & 	  	  & 	 3 	  & 	 987 	  & 	 1104 	  & 	 1018 	  & 	 806 	  & 	 783 	  & 	 182 	  & 	 40 	  & 	 - 	  & 	 17.03 	  & 	 0.64 	  & 	 4.46 	  \\
 \object{1730$-$130} 	  & 	 NRAO 530 	  & 	 LPQ 	  & 	 x 	  & 	 2 	  & 	 340 	  & 	 615 	  & 	 501 	  & 	 83 	  & 	 80 	  & 	 93 	  & 	 - 	  & 	 - 	  & 	 15.55 	  & 	 0.71 	  & 	 4.06 	  \\
 \object{1739$+$522} 	  & 	 S4 1739+52 	  & 	 HPQ 	  & 	 x 	  & 	 2 	  & 	 259 	  & 	 363 	  & 	 395 	  & 	 149 	  & 	 121 	  & 	 - 	  & 	 - 	  & 	 - 	  & 	 3.12 	  & 	 0.77 	  & 	 4.24 	  \\
 \object{1741$-$038} 	  & 	 PKS 1741-038 	  & 	 HPQ 	  & 	  	  & 	 2 	  & 	 - 	  & 	 222 	  & 	 299 	  & 	 272 	  & 	 296 	  & 	 111 	  & 	 - 	  & 	 - 	  & 	 8.93 	  & 	 0.70 	  & 	 3.03 	  \\
 \object{1749$+$096} 	  & 	 PKS 1749+096 	  & 	 BLO 	  & 	  	  & 	 5 	  & 	 695 	  & 	 1032 	  & 	 1088 	  & 	 583 	  & 	 465 	  & 	 154 	  & 	 - 	  & 	 51 	  & 	 10.09 	  & 	 0.85 	  & 	 0.94 	  \\
 \object{1921$-$293} 	  & 	  	  & 	 HPQ 	  & 	  	  & 	 1 	  & 	 401 	  & 	 694 	  & 	 657 	  & 	 - 	  & 	 - 	  & 	 120 	  & 	 - 	  & 	 45 	  & 	 $15.47^1$ 	  & 	 $0.69^1$ 	  & 	 $10.68^1$ 	  \\
 \object{2007$+$776} 	  & 	 S5 2007+77 	  & 	 BLO 	  & 	  	  & 	 1 	  & 	 232 	  & 	 - 	  & 	 436 	  & 	 92 	  & 	 84 	  & 	 - 	  & 	 - 	  & 	 - 	  & 	 2.97 	  & 	 0.67 	  & 	 3.34 	  \\
 \object{2145$+$067} 	  & 	  	  & 	 LPQ 	  & 	  	  & 	 4 	  & 	 304 	  & 	 821 	  & 	 818 	  & 	 551 	  & 	 496 	  & 	 134 	  & 	 - 	  & 	 34 	  & 	 11.77 	  & 	 0.41 	  & 	 3.12 	  \\
 \object{2200$+$420} 	  & 	 BL Lac 	  & 	 BLO 	  & 	 x 	  & 	 9 	  & 	 938 	  & 	 1115 	  & 	 1298 	  & 	 965 	  & 	 998 	  & 	 145 	  & 	 - 	  & 	 - 	  & 	 13.75 	  & 	 0.83 	  & 	 1.44 	  \\
 \object{2201$+$315} 	  & 	 4C 31.63 	  & 	 LPQ 	  & 	  	  & 	 2 	  & 	 - 	  & 	 - 	  & 	 - 	  & 	 280 	  & 	 251 	  & 	 - 	  & 	 - 	  & 	 - 	  & 	 6.38 	  & 	 0.62 	  & 	 4.55 	  \\
 \object{2223$-$052} 	  & 	 3C 446 	  & 	 BLO 	  & 	  	  & 	 3 	  & 	 532 	  & 	 770 	  & 	 837 	  & 	 237 	  & 	 240 	  & 	 145 	  & 	 - 	  & 	 39 	  & 	 10.86 	  & 	 0.63 	  & 	 5.84 	  \\
 \object{2227$-$088} 	  & 	  	  & 	 HPQ 	  & 	  	  & 	 1 	  & 	 - 	  & 	 - 	  & 	 - 	  & 	 33 	  & 	 27 	  & 	 20 	  & 	 - 	  & 	 14 	  & 	 3.03 	  & 	 0.74 	  & 	 3.56 	  \\
 \object{2230$+$114} 	  & 	 CTA 102 	  & 	 HPQ 	  & 	 x 	  & 	 1 	  & 	 - 	  & 	 983 	  & 	 909 	  & 	 295 	  & 	 293 	  & 	 106 	  & 	 - 	  & 	 - 	  & 	 7.57 	  & 	 0.62 	  & 	 2.23 	  \\
 \object{2251$+$158} 	  & 	 3C 454.3 	  & 	 HPQ 	  & 	 x 	  & 	 6 	  & 	 674 	  & 	 1070 	  & 	 1067 	  & 	 760 	  & 	 722 	  & 	 244 	  & 	 - 	  & 	 73 	  & 	 16.82 	  & 	 0.58 	  & 	 1.51 	  \\
 \object{2255$-$282} 	  & 	  	  & 	 LPQ 	  & 	  	  & 	 1 	  & 	 - 	  & 	 - 	  & 	 - 	  & 	 - 	  & 	 - 	  & 	 55 	  & 	 - 	  & 	 22 	  & 	 $10.14^1$ 	  & 	 $0.76^1$ 	  & 	 $5.77^1$ 	  \\
\hline 
\end{longtable}
\begin{list}{}{\setlength{\leftmargin}{45pt}}
\item \footnotesize{$^1$ = Values taken at 90\,GHz}
\end{list}
\end{landscape}
}

\end{document}